# Free Transverse Vibration Analysis of thin rectangular plates having arbitrarily varying non-homogeneity along two concurrent edge

Avinash Kumar

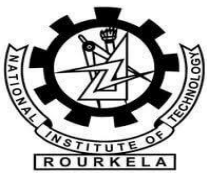

MECHANICAL ENGINEERING

National Institute of Technology, Rourkela

# Free Transverse Vibration Analysis of thin rectangular plates having arbitrarily varying non-homogeneity along two concurrent edge

*Thesis submitted in partial fulfilment of the requirements for the degree of*

**Master of Technology**

in

**Mechanical Engineering with Specialization in Machine Design and Analysis**

BY

**AVINASH KUMAR**

**Roll Number (216ME1349)**

based on research carried out under supervision of

**Prof. R.K BEHERA**

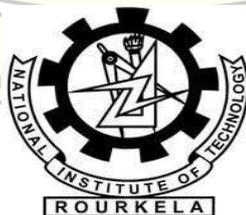

**May 2018**
**Department of Mechanical Engineering**
**National Institute of Technology Rourkela Odisha 769008**

I

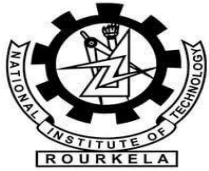

**MECHANICAL ENGINEERING**
**National Institute of Technology, Rourkela**

Prof. R.K Behera

Associate Professor                                                                                             May 24, 2018

# Supervisor's Certificate

This is to certify that the work presented in the thesis entitled **Free Transverse Vibration Analysis of thin rectangular plates having arbitrarily varying non-homogeneity along two concurrent edge** submitted by *AVINASH KUMAR*, **Roll Number 216ME1349**, is a record of original research carried out by him under my supervision and guidance in partial fulfilment of the requirements of the degree of *Master of Technology* in *MACHINE DESIGN AND ANALYSIS*. Neither this thesis nor any part of it has been submitted earlier for any degree or diploma to any institute or university in India or abroad.

\_\_\_\_\_\_\_\_\_\_\_\_\_\_\_\_\_\_\_\_\_\_\_\_\_\_\_\_

*Prof.  R.K BEHERA*

Dept. of Mechanical Engineering

National Institute of Technology, Rourkela

# Dedication

*Dedicated to my beloved family*

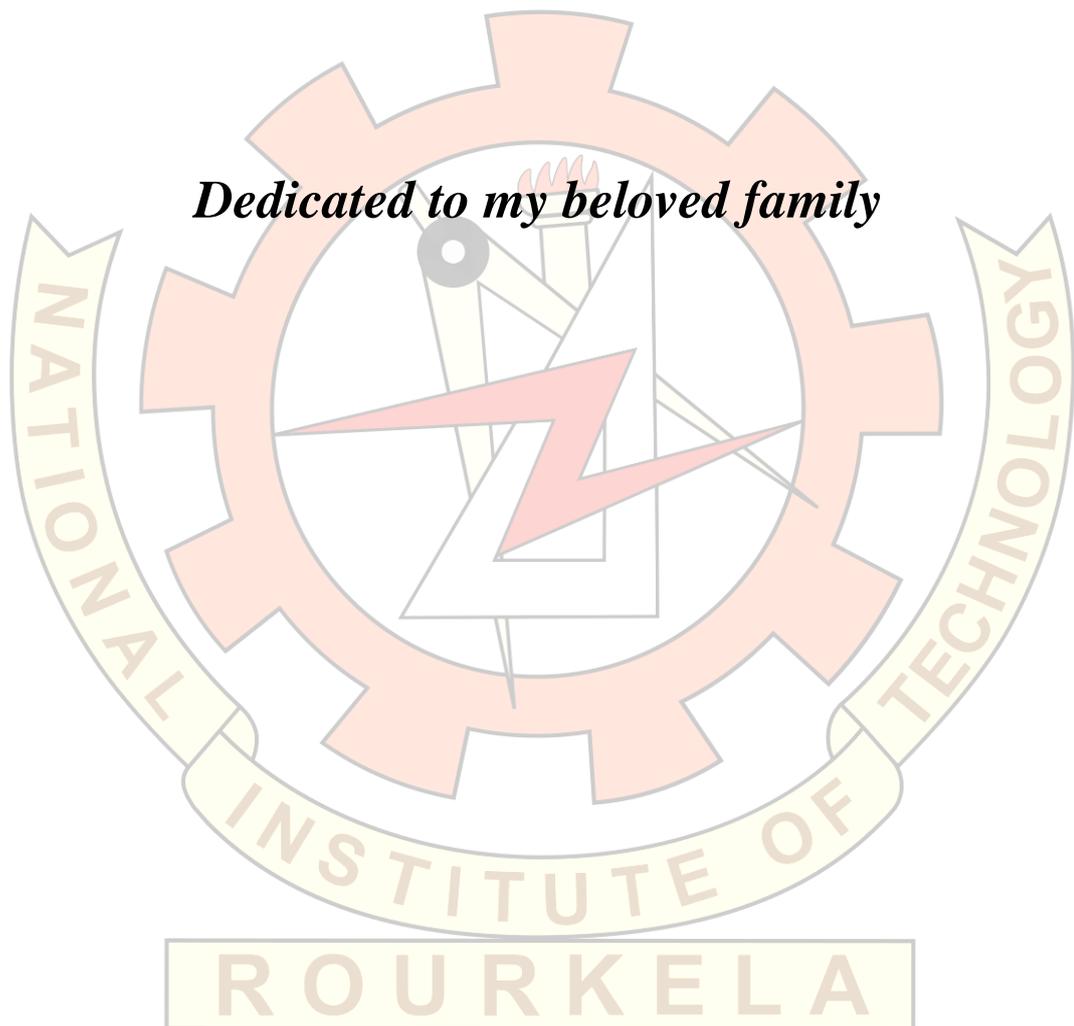

# Declaration of Originality

I, AVINASH KUMAR, Roll Number 216ME1349 hereby declare that this thesis entitled Free Transverse Vibration Analysis of thin rectangular plates having arbitrarily varying non-homogeneity along two concurrent edge presents my original work carried out as a Postgraduate student of NIT Rourkela and, to the best of my knowledge, contains no material previously published or written by another person, nor any material presented by me for the award of any degree or diploma of NIT Rourkela or any other institution. Any contribution made to this research by others, with whom I have worked at NIT Rourkela or elsewhere, is explicitly acknowledged in the dissertation. Works of other authors cited in this dissertation have been duly acknowledged under the sections "Reference" or "Bibliography". I have also submitted my original research records to the scrutiny committee for evaluation of my dissertation.

I am fully aware that in case of any non-compliance detected in future, the Senate of NIT Rourkela may withdraw the degree awarded to me on the basis of the present dissertation.

May 24, 2018                                                        Avinash Kumar
NIT Rourkela

# ACKNOWLEDGEMENTS

The two years I spent at the National Institute of Technology have been full of unforgettable memories. Thanks to many people who deserve my highest gratitude.

I would like to express our deep sense of gratitude to our supervisor **Prof. R.K BEHERA** for giving us this opportunity to work under him. Each and every discussion with him has been very educative. We are very thankful to him for the patience he had with us and for giving us absolute freedom in doing our task. We shall remain indebted to him for his constant support and guidance.

We are thankful to **Prof. D R PARHI**, Head of the Department, Department of MECHANICAL Engineering for providing us with necessary facilities for the research work.

I would also like to express my gratitude to PhD Scholar **KRISHANU GANGULY** and for his generous help for the completion of this project.

I would like to thank all my classmates and my friends especially. **BIREN KUMAR PRADHAN, SATYABRATA NAYAK and ABHISHEK PATRA.** A truly unbounded words of thanks to my family for their affection, constant encouragement and forbearance.

Finally, I bow myself to Almighty God whose blessings guard and guide me throughout my life.

May 24, 2018

NIT Rourkela                                                                                           Avinash Kumar

# ABSTRACT

In this paper, I presented the analysis and numerical results for free transverse vibration of thin rectangular plates having arbitrarily varying non-homogeneity with the in-plane coordinates along the two concurrent edges. For finding the general governing differential equation first used the Kirchhoff's plate theory by considering their assumptions.

After finding the governing equation for the plate. For the non- homogeneity a linear variation for Young's modulus and density of the plate has been assumed. Finite difference method (FDM) has been used to obtain the eigen value problem for such model plate for two different boundary condition at the edge namely (i) CCCC fully clamped (ii) CSCS two opposites are clamped and other two are simply supported.

By solving these eigen value problem using MATLAB, the lowest three eigen value have been reported as the first three natural frequencies for the three mode of vibration. The effect of various plate parameters such on the vibration characteristics has been analysed. Three dimensional mode shape has been plotted.

***Keywords: Non-homogeneity, flexural rigidity, FDM, Non-dimensional frequency parameters***

# CONTENTS

# List of Figure

# List of Table

# LIST OF ACRONYMS

| Symbol | Description |
|---|---|
| $x, y, z$ | Cartesian coordinate system |
| $X, Y, H, W$ | Non dimensional variables |
| $Q_x, Q_y$ | Transverse shearing force per unit length in the respective direction |
| $N_x, N_y, N_{xy}$ | In-plane Normal and Shear force per unit length |
| $M_x, M_y, M_{xy}$ | Bending moment and twisting moment intensities per unit length |
| $q(x, y)$ | Transverse external force per unit area |
| $w(x, y, t)$ | Displacement function in z-direction |
| $a, b, h$ | Length, width and thickness of the plate |
| $\rho(X, Y)$ | Mass density function |
| $\dfrac{\partial^2 w}{\partial t^2}$ | Acceleration in z-direction |
| $E(X, Y)$ | Young's modulus function |
| $\varepsilon_x, \varepsilon_y, \chi_{xy}$ | Normal and shear strain in the respective direction |
| $D = D(x, y)$ | Flexural rigidity function |
| $\vartheta$ | Poisson's ratio |
| $\lambda = a/b$ | Aspect ratio |
| $\alpha_1, \alpha_2$ | Non-homogeneity parameters |
| $\beta_1, \beta_2$ | Density parameters |
| $E_0, \rho_0$ | Young's modulus and density of the plate material at $X = 0$ and $Y = 0$ |
| $p, q, r, s$ | Positive integers |
| $N, M$ | Number of grid points in x any y direction |
| $\omega$ | Circular frequency |
| $\Omega$ | Non- dimensional frequency parameters |
| $C$ | Clamped |
| $S$ | Simply supported |
| $\Delta X, \Delta Y$ | Equally spaced grid length and width |
| $W(i, j)$ | Non dimensional displacement at $(i, j)$ points |
| LVN | Linear Variation Non-homogeneity |
| PVN | Parabolic Variation Non-homogeneity |

# Chapter-1

# INTRODUCTION

# 1. Introduction

**Plate structure**

Plate structure element is characterised by its property such as, it is a three dimensional solid whose thickness is very small as compared to the other dimensions and the effects of the load that applied on it generate the stresses which are generally normal stress only. The plate bounded by the two parallel planes, and the distance between the plate is called the thickness of the plate. The 2D plate theory give the approximate results such as 3D plate like structure. Because the one dimension of the plate is very less compared to other two dimension of the plate. So it is a plain stress condition in which the stresses like $\sigma_{zz} = \sigma_{zx} = \sigma_{zy} = 0$. Along with the stresses the shear strain $\chi_{yz} = \chi_{xz}$ is also zero.

## 1.2 Background

Several years ago, in the aerospace field an engineer needed to know the first three frequency and mode shape of a rectangular plate of a certain aspect ratio along with the certain boundary condition. After various literature search for several weeks, they only get two first frequency but not get any accurate mode shape. Since he had neither the analytical capability of solving the problem nor the money and time needed for experimental programme.

Since we know that, as the technology is increasing rapidly, the requirement of the advanced material is also at a high demand due to their extensive use in the engineering applications, especially for the performance in the high temperature environment. Because these material are stiffer, stronger and corrosion resistance. This literature provides free vibration analysis of thin rectangular isotropic plate having arbitrarily varying non-homogeneity with the in-plane coordinates along the two concurrent edges.

But now-a-days, technologists are able to tailor advanced material by mixing two or more materials to get the desired properties along the one/more direction due to their extensive demand in the many field of modern engineering applications, and especially for performance in the high temperature environment. Usually, these materials are stiffer, stronger and corrosion resistant than the other conventional material used earlier. The structural element made up of such material will be non-homogeneous by nature in which the material properties may vary continuously in a certain manner either along a line or in a plane or in a space. During the past few decades, due to development of high speed digital computer and numerical methodologies a huge amount of work analysing the dynamic behaviour of the plates of various geometries with different boundary condition. Despite the aforementioned researches on the vibration of the non- homogeneous rectangular plates, there exists very few researches in which two-dimensional (2D)- arbitrary variation for the non-homogeneity of the plate material has been considered.

## 1.3 Present Work

The present work is done to fill this gap of isotropic plate with the uniform thickness. For such plate model, the governing differential equation of motion has been solved employing the Finite Difference Method (FDM). Further, this consideration for non-homogeneity together with aspect ratio of the plate shows their effect on the natural frequencies for the all possible combination of the classical boundary condition at the four edge of the plate material.

But, in this paper linear model for non-homogeneity of the isotropic material have been considered assuming that the Young's modulus and density are varying in same distinct manner along the both in plane coordinates. The non-homogeneity arises due to arbitrary variation in Young's modulus and density. All of these investigation have been taking poison's ratio $V$ as constant. This type of non-homogeneity arises during the fibre reinforced plastic structure, which use fibre of different strength properties along two mutually perpendicular directions along the edges of the plate.

Since we know that there are totally 21 possible boundary conditions for a rectangular plate. this will generate a huge data therefore in this present work, the numerical computation has been carried out only for the two boundary condition, namely (i) CCCC- all the four edge are clamped, (ii) CSCS- two opposite edges are simply supported and the other opposite edges are clamped with the linear variation of Young's modulus and the density. The effect of non-homogeneity parameters, density parameters and the aspect ratio on the natural frequencies has been investigated for the first three modes of vibration.

## 1.4 Novelties of this paper

(i) Plate type structure of 2D model for the non-homogeneity has been proposed.

(ii) The Young's modulus and density of the plate material are assumed as a function of in-plane coordinates x and y.

(iii) The proposed model encompasses various unidirectional/bidirectional, linear and non-linear models.

(iv) FDM solution has been obtained successfully to find the mode shape of this model plate structure Natural frequency found by using the Rayleigh method with deflection function as a product of beam function.

(v) This analysis may be useful for missiles and aircraft designers, solid state physicist and, in general people engaged in material science.

# Chapter 2

# Literature Review

## 2.1. Literature Review

Since we know that, as the technology is increasing rapidly, the requirement of the advanced material is also at a high demand due to their extensive use in the engineering applications, especially for the performance in the high temperature environment. Because these material are stiffer, stronger and corrosion resistance. This literature provides free vibration analysis of thin rectangular isotropic plate having arbitrarily varying non-homogeneity with the in-plane coordinates along the two concurrent edges.

The study of non-homogeneous material with fair amount of accuracy is of practical importance from the view point of design engineers. In this regard, the earliest model for the non-homogeneity was proposed by Bose [1] in which Young's modulus and density of the plate material are supposed to vary with the radius vector i.e. $(E, \rho) = (E_0, \rho_0)$r. An approximate solution is obtained using Ritz method. The frequency parameter $\Omega$ is found to increase with the increase in non-homogeneity parameter as well as taper parameter, while it decreases with the increase in density parameter.

Later on, Biswas [2] developed a model for non-homogeneity considering the exponential variation for torsional rigidity $\mu = \mu_0 e^{-\mu_1 z}$ and the material density $\rho = \rho_0 e^{\mu_1 z}$, where $\mu_0, \rho_0$ and $\mu_1$ are constants. DQM (Differential Quadrature Method) has been used to analyse the free vibration of non- homogeneous orthotropic rectangular plate of parabolic varying thickness resting on Winkler foundation.

Das and Mishra [3] deals with the transverse vibration of elliptical and circular plate using two dimensionally boundary characteristics orthogonal polynomial using Rayleigh Ritz method by considered the non-homogeneity of the plate characterized by taking Young's modulus $E = E_0(1-\rho^z)^\alpha$ and density $\sigma = \sigma_0(1-\rho^z)^{\alpha-2}$, $0 \leq \rho \leq 1$ and $\alpha$ is an integer greater than 3.

Rao et al. [4] proposed a model for non-homogeneous isotropic thin plate by assuming linear variation for Young's modulus and density given by $E = E_0(1+\alpha x)$ and $\rho = \rho_0(1+\beta x)\alpha$, $\beta$ being constant.

The free transverse vibration of non -homogeneous circular plate of linearly varying thickness on the basis of classical plate theory has been analysed with help of Tomar JS, Gupta DC and Jain NC [5]. In which The frequency parameter increases with increase the rate of increase for free plate is higher than simply-supported plate but lesser than clamped plate.

The Natural frequencies for free vibration of non-homogeneous elliptic and circular plates using two-dimensional orthogonal polynomials has been analyse with help of Chakravety and Pety [6]. In this paper the non- homogeneity is taken by considering linear variation of the Young's Modulus and parabolic variation of the density of the material. The first five natural frequencies have been analysed in this paper.

Effect of non-homogeneity on asymmetric vibrations of non-uniform circular plates has been analysed by Seema Sharma, R. Lal, Neelam Singh [7] where the non-homogeneity arises due to exponential variation of the Young's modulus as well as density parameters along the radial direction. The first three natural frequencies are found with using of Ritz method. The effect of the taper parameters and the density parameters has been analysed in this paper.

Free transverse vibrations of nonhomogeneous orthotropic rectangular plates with bilinear thickness variation resting on Winkler foundation are presented by Yazuvendra kumar, R. Lal [8] by using two dimensional boundary characteristics orthogonal polynomial using Rayleigh Ritz method on the basis of classical plate theory. Orthogonal polynomial has been generated by using Gram-Schmidt. The non-homogeneity is considering by taking the linear variation of the Young's modulus as well as density. The two dimensional thickness variation is considered by taken Cartesian product of linear variation. In the previous all literature papers there is assumption taken that the Poisson's ratio is constant throughout. However, in the two studies Rossi and Laura [9] it has been found that effect of the Poisson's ratio on the frequency is not appreciable and reported as considerably less than 1%.

Free vibration analysis of clamped orthotropic plate material using Lagrangian multiplier technique by R. L Ram Kumar et al., [10]. Transverse shear deformation effects were included using a higher-order plate theory. The validated free vibration analysis has also been used to develop an analysis to predict the dynamic response of clamped orthotropic plates subjected to low-velocity impact by a hard object. Effect of non-homogeneity parameters, aspect ratio and thickness variation together with foundation parameter on the natural frequencies has been illustrated for the first three modes of vibration for four different combinations of clamped, simply supported and free edges correct to four decimal places.

The hierarchical finite element method is used to determine the natural frequencies and modes of a flat, rectangular plate by Bardell [11] Ten different boundary conditions—including free edges and point supports—are considered in this paper. Extensive results are presented for each case (including the variation of frequency with the aspect ratio and the Poisson ratio), and these are shown to be in very good agreement with the work of other investigators. This confirms both the applicability and accuracy of solution of the HFEM to problem of this type.

Kantorovich method has also been used to find the natural frequencies of orthotropic rectangular plates Sakata et al., [12]. In this method the natural frequencies are obtained by the successive reduction of the plate partial differential equation and by solving the results. The reduction of the partial differential equation is obtained by assuming an approximate solution satisfying the boundary condition along the one direction and then apply the Kantorovich method.

Vibration analysis of laminated composite plates on elastic foundations by Ishan kucukrendeci and Hasan Kucuk [13]. In this study the composite plates consist of layers and layer are arranged at different angles. And this plate supported by the springs of five different points. The finite element method is used to find the natural frequencies of this plate model. The boundary condition of the plate having three different conditions i.e clamped, simply supported and free edge. MATLAB is used to solve the equation.

Free vibration analysis of rectangular plates with variable thickness M. Huang [14]. In this paper an approximate method which is based on the Green function of a ate plate is used. The Green function of a rectangular plate with arbitrary variable thickness is obtained as a discrete form of solution for the deflection of the plate with concentrated load. The solution is obtained at each discrete point which are equally distributed on the plate. After applying the green function, the free vibration of the plate converted into the eigen value problem of matrix. The convergence an accuracy of the numerical solution for the natural frequencies parameters calculated by the proposed method.

Hybrid method for the vibration analysis of the rectangular plate by Kerboua et al., [15]. This paper basically related to the dynamic analyses of the rectangular plates. Mathematical model is developed by the hybrid method and finite element method and Sander's shell theory is used to find out the frequency of this plate material. The in plane displacement membrane component is modelled by bilinear polynomials and the out of plane, normal to the mid plane displacement component is modelled by an exponential function. the displacement function is obtained by the exact solution of the equilibrium equation and the mass and the stiffness matrices are then determined by exact analytical integration to established the dynamic plate equation.

Vibration analysis of the structure with Differential quadrature approach by Richard Bellman et al., [16]. This technique could be utilized in a simple manner to obtain the solution of non-homogeneous differential equations derived from applications of the theory of invariant imbedding to transport process. In this paper the ideas offer a new technique for the numerical solution of initial value problem for the ordinary and partial differential equations for a certain complex problem

A general boundary condition new approach for implementing the general boundary condition in the GDQ free vibration analysis of rectangular plates is presented in this paper by C. Shu, H. Du [17]. In this approach boundary condition is directly coupled with the governing equation. The accuracy of this approach are demonstrated through its application to the vibration analysis of a rectangular plate with various combination of free edges and corners. In this approach the equally spaced grid distribution method has been used to find the natural frequencies of the plate.

Harmonic differential quadrature method and its application to analyse of structural components String et al., [18]. This method can be used to analysed the buckling and the free vibration of the beams and rectangular plates. In this method a new technique is used to find out the weightage coefficient for the differential quadrature. It is more efficient than the DQ method especially for the higher order frequencies and the buckling load of the rectangular plates under the wide range of the aspects ratio.

Kuldeep S. Virdi [19] Finite difference method for nonlinear analysis of Structures. This paper describes a sustained line of development of methods of nonlinear analysis of structural element and frames using FDM. This method is also extended to use in composite frames and structure. A good correlation with this experiment has been obtained for the deflection. Also the problem of nonlinear response of the structure has been successfully solved.

Analysis of beam by the FDM (Finite Difference Method) Gautam Chattopadhyay [20]. In this paper studied how the FDM technique can be used to find the deflection and frequency of the beam at each node of the distributed grid points.

Accuracy study of the FDM by Nancy Jane Cyrus, Robert E. Fulton [21]. In this paper FDM used for linear differential equation. Definite error for each expression is derived from the Taylor series. This method is used to assess the accuracy of two alternate forms of central finite difference approximations for solving the boundary value problem of the structural analysis which are governed by the certain equations containing variable coefficient.

# Chapter 3

# MATHEMATICAL FORMULATION

## 3.1. Mathematical Formulation

For the mathematical formulation, considered a non-homogeneous isotropic rectangular plate of uniform thickness $h$, density $\rho(x, y)$ having a domain $0 \leq x \leq a, 0 \leq y \leq b$, where $a$ and $b$ are length and breadth of the plate respectively. The x-axis and y-axis are taken along the edge of the plate and the z-axis is perpendicular to the xy- plane. The middle surface being $z = 0$ and the origin is at one of the corners of the plate as shown in figure 2(a).

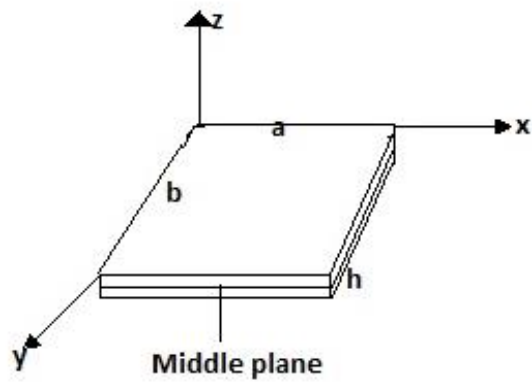

Figure 1 Geometric of the isotropic plate

### 3.1.1 Kirchhoff- love plate theory

To find the equation of motion of this plate model. Let us taken the infinitesimal plate of length, breadth and height dx, dy and h respectively. The Kirchhoff–Love theory of plates is a two-dimensional mathematical model that is used to determine the stresses and deformation in thin plates subjected to forces and moments. It is extension of Euler-Bernoulli beam theory developed by Love in 1988.

### 3.1.2 Assumption

The assumptions which is used to find the equation of motion of this plate model are-

(i) straight lines normal to the mid-surface remain straight after deformation.
(ii) straight lines normal to the mid-surface remain normal to the mid-surface after deformation.
(iii) the thickness of the plate does not change during a deformation.

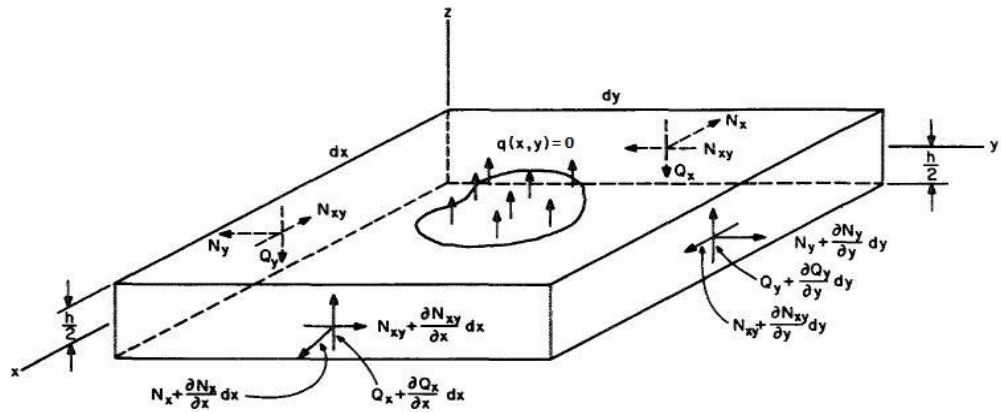

Figure 2 Force (intensity) acting on the plate element

The Figure 2 shows transverse shearing force intensities $Q_z$ and $Q_y$, the in-plane normal and shearing force intensities $N_x$, $N_y$ and $N_{xy}$, and their incremental changes are shown acting on the sides of the element, with positive forces acting in positive directions on positive faces. The shearing force $N_{xy}$ are identical on the face at x = 0 and y = 0. Because the shearing stress causing them are equal.

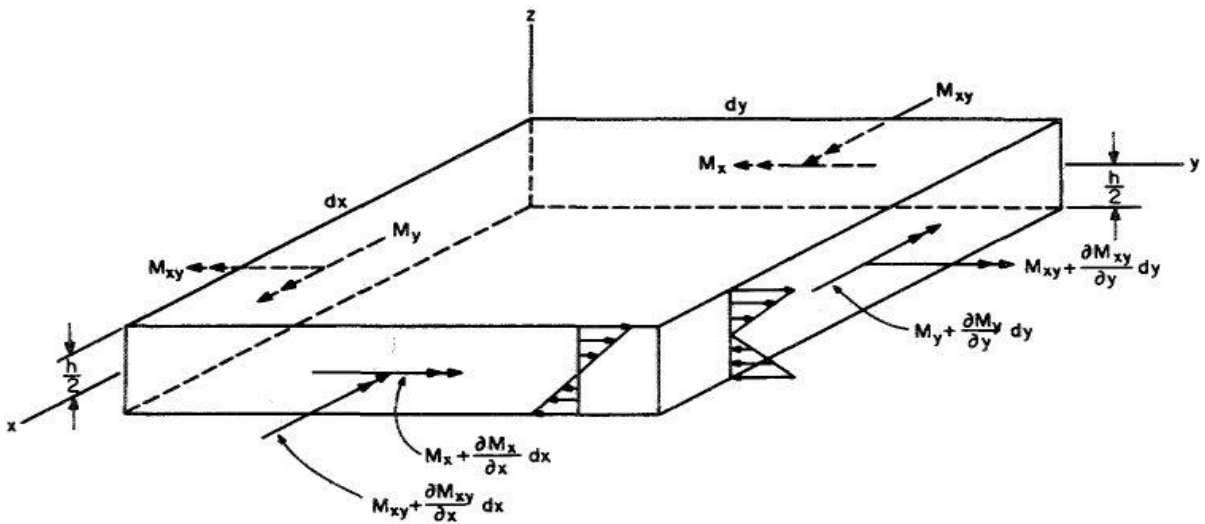

Figure 3 Moment (intensity) acting on the plate element

Figure 3 shows the bending moment and the twisting moment intensities and their incremental change. The twisting moment $M_{xy}$ are identical on the face at x = 0 and at y = 0. Because the shearing stress causing them are equal

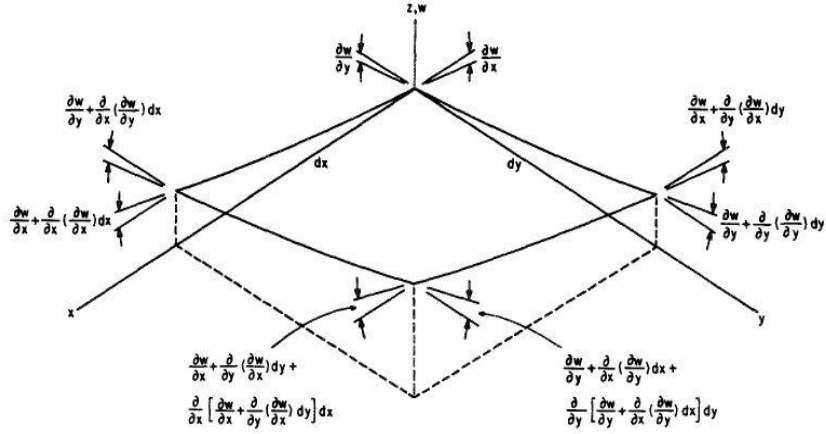

Figure 4 shows the deformed middle surface of the plate element showing slopes and their change

The figure 4 shows that incremental changes are shown at all the corner of the element with positive changes assume in positive direction. For small displacement, it is assumed that the slope and the sine of the angle are equivalent.

For equilibrium summing the forces in z-direction

$$\frac{\partial Q_x}{\partial x}dxdy + \frac{\partial Q_y}{\partial y}dydx - N_x dy \frac{\partial w}{\partial x} + \left(N_x + \frac{\partial N_x}{\partial x}dx\right)dy$$
$$\left(\frac{\partial w}{\partial x} + \frac{\partial^2 w}{\partial x^2}dx\right) - N_y dx \frac{\partial w}{\partial y} + \left(N_y + \frac{\partial N_y}{\partial y}dy\right)dx\left(\frac{\partial w}{\partial y} + \frac{\partial^2 w}{\partial y^2}dy\right)$$
$$- N_{xy} dx \frac{\partial w}{\partial x} + \left(N_{xy} + \frac{\partial N_{xy}}{\partial y}dy\right)dx\left(\frac{\partial w}{\partial x} + \frac{\partial^2 w}{\partial x \partial y}dy\right) -$$
$$N_{xy} dy \frac{\partial w}{\partial y} + \left(N_{xy} + \frac{\partial N_{xy}}{\partial x}dx\right)dy\left(\frac{\partial w}{\partial y} + \frac{\partial^2 w}{\partial x \partial y}dx\right) +$$
$$qdxdy = \rho dxdy \frac{\partial^2 w}{\partial t^2}$$

(3.1)

We know that equilibrium equation from theory of elasticity by neglecting the inertial force and transverse shear stress. because $\tau_{yz}$ and $\tau_{zx}$ are small relative to other stress.

$$\frac{\partial \sigma_x}{\partial x} + \frac{\partial \tau_{xy}}{\partial y} = 0$$
$$\frac{\partial \tau_{xy}}{\partial x} + \frac{\partial \sigma_y}{\partial y} = 0$$

(3.2)

Because these equations must be satisfied for every infinitesimal thickness dz of the plate element, their integral over the thickness also satisfied

$$\frac{\partial N_x}{\partial x} + \frac{\partial N_{xy}}{\partial y} = 0$$

$$\frac{\partial N_{xy}}{\partial x} + \frac{\partial N_y}{\partial y} = 0 \qquad (3.3)$$

Using equation (2.3) in equation (2.1), the equilibrium equation in z-direction

$$\frac{\partial Q_x}{\partial x} + \frac{\partial Q_y}{\partial y} + N_x \frac{\partial^2 w}{\partial x^2} + N_y \frac{\partial^2 w}{\partial y^2} + 2N_{xy} \frac{\partial^2 w}{\partial x \partial y} + q$$

$$= \rho h \frac{\partial^2 w}{\partial t^2} \qquad (3.4)$$

Similarly, the forces in x and y-direction

$$\frac{\partial N_x}{\partial x} + \frac{\partial N_{xy}}{\partial y} - \frac{\partial}{\partial x}\left(Q_x \frac{\partial w}{\partial x}\right) - \frac{\partial}{\partial y}\left(Q_y \frac{\partial w}{\partial y}\right) = \rho \frac{\partial^2 u}{\partial t^2}$$

$$\frac{\partial N_{xy}}{\partial x} + \frac{\partial N_y}{\partial y} - \frac{\partial}{\partial x}\left(Q_x \frac{\partial w}{\partial y}\right) - \frac{\partial}{\partial y}\left(Q_y \frac{\partial w}{\partial y}\right) = \rho \frac{\partial^2 v}{\partial t^2} \qquad (3.5)$$

Now, moment about x and y-axis are

$$(Q_x) + \frac{\partial M_x}{\partial x} - \frac{\partial M_{xy}}{\partial y} = \frac{\rho h^3}{12} \frac{\partial^3 w}{\partial x \partial t^2}$$

$$(Q_y) + \frac{\partial M_{xy}}{\partial x} - \frac{\partial M_y}{\partial y} = \frac{\rho h^3}{12} \frac{\partial^3 w}{\partial y \partial t^2}$$

$$(3.6)$$

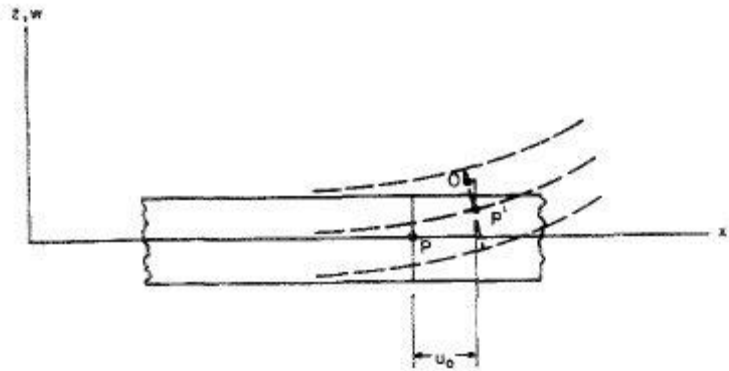

Figure 5 Kinematics of plate deformation

An edge view of the portion of the plate material shown in figure. The deformed shape is shown in broken lines while the un-deformed shape is shown in unbroken lines. The longitudinal elastic displacement of a point P on the mid-plane is depicted by $u_0$. The longitudinal displacement of the points within the plate written as

$$u = u_0 - z\frac{\partial w}{\partial x}$$
$$v = v_0 - z\frac{\partial w}{\partial y}$$
(3.7)

The strain displacement function is

$$\varepsilon_x = \frac{\partial u}{\partial x} = \frac{\partial u_0}{\partial x} - z\frac{\partial^2 w}{\partial x^2}$$
$$\varepsilon_y = \frac{\partial v}{\partial y} = \frac{\partial v_0}{\partial y} - z\frac{\partial^2 w}{\partial y^2}$$
$$\chi_{xy} = \frac{\partial v}{\partial x} + \frac{\partial u}{\partial y} = \left(\frac{\partial v_0}{\partial x} + \frac{\partial u_0}{\partial y}\right) - 2z\frac{\partial^2 w}{\partial x \partial y}$$
(3.8)

Since we know that for the isotropic material, strain stress relationship is

$$\varepsilon_x = \frac{1}{E}(\sigma_x - \vartheta\sigma_y)$$
$$\varepsilon_y = \frac{1}{E}(\sigma_y - \vartheta\sigma_x)$$
$$\chi_{xy} = \frac{2(1+\vartheta)}{E}\tau_{xy}$$
(3.9)

$$N_x = \int_{-h/2}^{h/2} \sigma_x dz, \quad M_x = \int_{-h/2}^{h/2} \sigma_x z dz,$$
$$N_y = \int_{-h/2}^{h/2} \sigma_y dz, \quad M_y = \int_{-h/2}^{h/2} \sigma_y z dz,$$
$$N_{xy} = \int_{-h/2}^{h/2} \tau_{xy} dz, \quad M_{xy} = \int_{-h/2}^{h/2} \tau_{xy} z dz$$

(3.10)

Now the force and moment integral for the homogeneous plate is-

With the help of equation, (2.8) and (2.9) get the value of force intensities and moment intensities.

$$M_x = -D\left(\frac{\partial^2 w}{\partial x^2} + \vartheta \frac{\partial^2 w}{\partial y^2}\right),$$

$$My = -D\left(\frac{\partial^2 w}{\partial y^2} + \vartheta \frac{\partial^2 w}{\partial x^2}\right),$$

$$M_{xy} = -D(1-\vartheta)\left(\frac{\partial^2 w}{\partial x \partial y}\right)$$

(3.11)

Now using the equation (2.11) in equation (2.6) to find the $Q_x$ and $Q_y$

$$Q_x = -\frac{\partial D}{\partial x}\left(\frac{\partial^2 w}{\partial x^2} + \vartheta \frac{\partial^2 w}{\partial y^2}\right) - D\left(\frac{\partial^3 w}{\partial x^3} + \frac{\partial^3 w}{\partial x \partial y^2}\right) - (1-\vartheta)\frac{\partial D}{\partial y}\frac{\partial^2 w}{\partial x \partial y}$$

$$Q_y = -\frac{\partial D}{\partial y}\left(\frac{\partial^2 w}{\partial y^2} + \vartheta \frac{\partial^2 w}{\partial x^2}\right) - D\left(\frac{\partial^3 w}{\partial y^3} + \frac{\partial^3 w}{\partial x^2 \partial y}\right) - (1-\vartheta)\frac{\partial D}{\partial x}\frac{\partial^2 w}{\partial x \partial y}$$

(3.12)

Now put the value of $Q_x, Q_y$ in the equation (2.1), and the in-plane forces are first determined from the equation (2.10). They are uncoupled each other in case of bending and stretching. Since the analysis of this plate model is based on free vibration i.e. $(q=0)$. Therefore the equilibrium equation (2.1) will become

$$\nabla^2\left(D\nabla^2 w\right) - (1-\vartheta)$$
$$\left[\frac{\partial^2 D}{\partial x^2}\frac{\partial^2 w}{\partial y^2} - 2\left(\frac{\partial^2 D}{\partial x \partial y}\right)\left(\frac{\partial^2 w}{\partial x \partial y}\right) + \frac{\partial^2 D}{\partial y^2}\frac{\partial^2 w}{\partial x^2}\right]$$
$$+\rho h \frac{\partial^2 w}{\partial t^2} = 0 \tag{3.13}$$

Where,

$$D = D(x,y) = \frac{E(x,y)h^3}{12(1-\vartheta^2)}$$

### 3.1.3 Consideration of Non-homogeneity

For the harmonic motion the displacement w is assumed to be

$$w(x,y,t) = \overline{w}(x,y)e^{i\omega t} \qquad (3.14)$$

Now for considering the non-homogeneity considering the non-dimensional variables $X = x/a$, $Y = y/b$, $H = h/a$ and $W = \overline{w}/a$. And the variation of the Young's modulus and density of the plate material as a function of in-plane coordinate.

$$\begin{aligned} E(X,Y) &= E_0\left(1 + \alpha_1 X^p + \alpha_2 Y^q\right), \\ \rho(X,Y) &= \rho_0\left(1 + \beta_1 X^r + \beta_2 Y^s\right) \end{aligned} \qquad (3.15)$$

Thereafter the equation (2.13) reduces to

$$(1 + \alpha_1 X^p + \alpha_2 Y^q)\left(\frac{\partial^4 W}{\partial X^4} + 2\lambda^2 \frac{\partial^4 W}{\partial X^2 \partial Y^2} + \lambda^4 \frac{\partial^4 W}{\partial Y^4}\right) + 2p\alpha_1 X^{p-1}\left(\frac{\partial^3 W}{\partial X^3} + \lambda^2 \frac{\partial^3 W}{\partial X \partial Y^2}\right)$$

$$+ 2q\alpha_2 Y^{q-1}\left(\lambda^4 \frac{\partial^3 W}{\partial Y^3} + \lambda^2 \frac{\partial^3 W}{\partial X^2 \partial Y}\right) + p(p-1)\alpha_1 X^{p-2}\left(\frac{\partial^2 W}{\partial X^2} + \nu\lambda^2 \frac{\partial^2 W}{\partial Y^2}\right)$$

$$+ q(q-1)\alpha_2 Y^{q-2}\left(\nu\lambda^2 \frac{\partial^2 W}{\partial X^2} + \lambda^4 \frac{\partial^2 W}{\partial Y^2}\right) - \Omega^2\left(1 + \beta_1 X^r + \beta_2 Y^s\right)W = 0 \qquad (3.16)$$

The consideration of this type of non-homogeneity help in complex problem and be practical use for Geologist to apply it as a tool in examining the heterogeneity of the material.

## 3.2 Boundary Condition

The two type of boundary condition namely, CCCC and CSCS have been considered where the symbol C, S denotes the clamped and the simply supported. In a notation such as CSCS the first symbol indicates the condition at X= 0, the second is at Y= 0, the third at X= 1, the fourth at Y= 1.

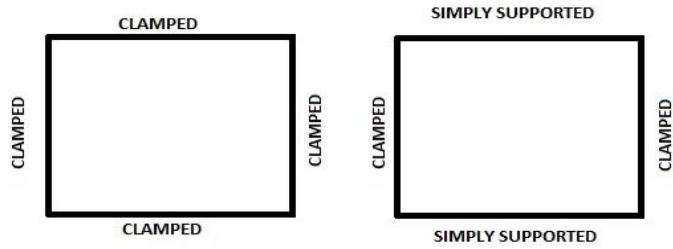

Figure 6 Boundary condition

### 3.2.1 Clamped edge

$$W = 0, \frac{dW}{dX} = 0 \text{ at } X = 0 \text{ and } 1$$
$$W = 0, \frac{dW}{dY} = 0 \text{ at } Y = 0 \text{ and } 1$$

(3.17)

### 3.2.2 Simply Supported edge

$$W = 0, \frac{d^2W}{dX^2} = 0 \text{ at } X = 0 \text{ and } 1$$
$$W = 0, \frac{d^2W}{dY^2} = 0 \text{ at } Y = 0 \text{ and } 1$$

(3.18)

# Chapter 4

# METHOD OF SOLUTION

## 4.1. Method of Solution

All together there are 21 combinations of simple boundary conditions (i.e., either clamped (C), simply supported (SS), or free (F)) for rectangular plates. Frequency parameter does not depend upon the Poisson's ratio until at least one of the edge of the plate is free. However because $D$ contains $\vartheta$, the frequency themselves depends upon $\vartheta$ for all cases.

### 4.1.1. Method of Solution Using FDM Technique

Finite difference method is important tool for the analysis of structures. Due to FDM technique a sustained line of development of methods of nonlinear analysis of structural element and frames. This method is also extended to use in composite frames and structure. A good correlation with this experiment has been obtained for the deflection. Also the problem of nonlinear response of the structure has been successfully solved.

Since we know that the FDM technique is a numerical method to find the approximate solution of a differential equation. It is very helpful to find the solution of partial differential equation. In this method at first the problem has to discretized. This is done by dividing the problem domain into a uniform grid. Therefore, it is also known as discretization method.

For this plate model divided the model into $M \times N$ grid points. The length and the width of each grid is $\Delta X$ and $\Delta Y$ respectively.

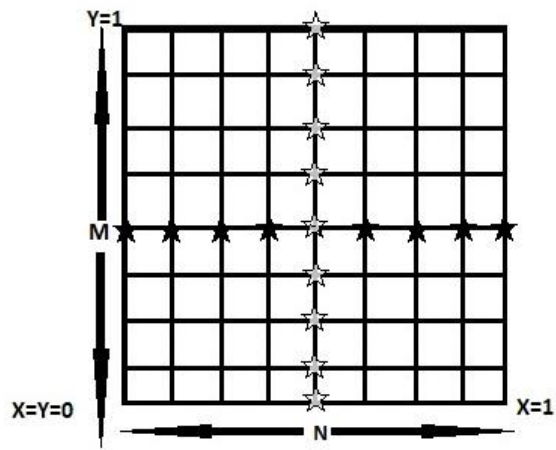

Figure 7 Grid point distribution on a rectangular domain

For applying the method FDM, one has to divide the computational domain of the plate $\{0 \leq X \leq 1, 0 \leq Y \leq 1\}$ by drawing the line parallel to $X$-axis and $Y$ - axis respectively. The total number of functional value in the whole domain is $N \times M$. Equally space grid points have been considered for the analysis of the plate model.

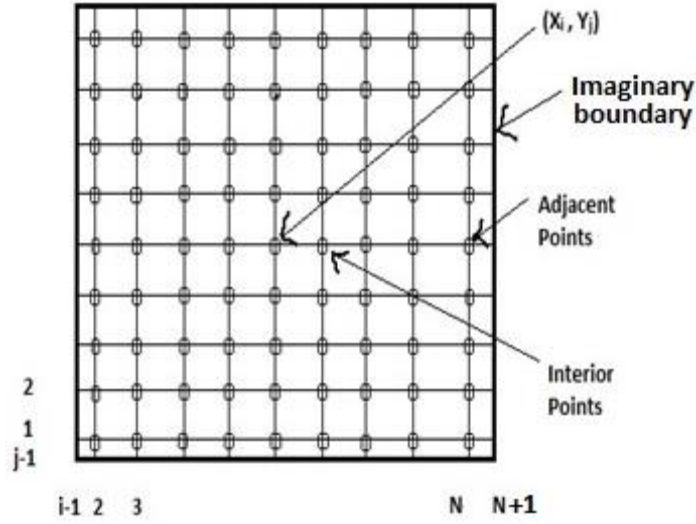

Figure 8 Demonstration of interior points, adjacent points and boundary of a rectangular plates

Figure (8) shows that number of grid points vary from $(i-1)$ to $N$ in x-direction and $(j-1)$ to $M$ in the y-direction. Let choose a grid of coordinate $(X_i, Y_j)$ and displacement function at that coordinate will be $W(i,j)$. The equation (3.16) have the 4$^{th}$ order partial differential term of two variables. Now for solving the equation (3.16) by FDM technique we have to find the value of partial differential part in the form of forward difference.

e.g.

$$\frac{\partial^3 W}{\partial X^3} = \frac{W(i+2, j) - W(i-2, j) - 2\{W(i+1, j) - W(i-1, j)\}}{\Delta X^3},$$
$$\frac{\partial^3 W}{\partial Y^3} = \frac{W(i, j+2) - W(i, j-2) - 2\{W(i, j+1) - W(i, j-1)\}}{\Delta Y^3} \quad (4.1)$$

$$\frac{\partial^3 W}{\partial X \partial Y^2} = \frac{W(i+1, j+1) + W(i+1, j-1) - \Gamma_1 - W(i-1, j+1) - W(i-1, j-1)}{\Delta X \Delta Y^2}$$
$$\frac{\partial^3 W}{\partial X^2 \partial Y} = \frac{W(i+1, j+1) + W(i-1, j+1) - \Gamma_2 - W(i+1, j-1) - W(i-1, j-1)}{\Delta X^2 \Delta Y}$$

(4.2)

$$\frac{\partial^4 W}{\partial X^2 \partial Y^2} = \frac{W(i+1, j+1) + W(i+1, j-1) - \Gamma_1 - \Gamma_2 + W(i-1, j+1) + W(i-1, j-1) + 4W(i,j)}{\Delta X^2 \Delta Y^2}$$
$$\Gamma_1 = 2\{W(i+1, j) - W(i-1, j)\}$$
$$\Gamma_2 = 2\{W(i, j+1) - W(i, j-1)\}$$

$$\frac{\partial^4 W}{\partial X^4} = \frac{W(i+2,j) - W(i-2,j) - 4\{W(i+1,j) - W(i-1,j)\} + 6W(i,j)}{\Delta X^4}$$

$$\frac{\partial^4 W}{\partial Y^4} = \frac{W(i,j+2) - W(i,j-2) - 4\{W(i,j+1) - W(i,j-1)\} + 6W(i,j)}{\Delta Y^4} \qquad (4.3)$$

Now put the required above value in the equation (2.16) we get

$$W(i,j) = -\frac{R}{S} \qquad (4.4)$$

Where,

$$\begin{aligned}
R &= \Delta_1(R_1) + \Delta_2(R_2) + \Delta_3(R_3) + \Delta_4(R_4) + \Delta_5(R_5) \\
S &= \Delta_1(S_1) - S_2 + \Delta_5 S_3 + \Delta_6 \\
R_1 &= (\lambda^4 \Psi_1 + 2\lambda^2 \Psi_2 + \Psi_3), \\
R_2 &= (\Psi_4 + \lambda^2 \Psi_5), \; R_3 = (\lambda^4 \Psi_6 + \lambda^2 \Psi_7) \\
R_4 &= (\Phi_1 + \vartheta \lambda^2 \Phi_2), \; R_5 = (\vartheta \lambda^2 \Phi_1 + \lambda^4 \Phi_3) \\
S_1 &= \left(\frac{6}{\Delta X^4} + \frac{8\lambda^2}{\Delta X^2 \Delta Y^2} + \frac{6\lambda^2}{\Delta Y^4}\right), \; S_2 = \left(\frac{2\Delta_4}{\Delta X^2} + \frac{2\vartheta \lambda^2}{\Delta Y^2} + \right) \\
S_3 &= \left(\frac{2\Delta_5 \vartheta \lambda^2}{\Delta X^2}\right), \; S_4 = \left(\frac{6\Delta_5 \lambda^4}{\Delta Y^4}\right)
\end{aligned} \qquad (4.5)$$

$$\Delta_1 = \left(1 + \alpha_1 X^p + \alpha_2 Y^q\right),$$
$$\Delta_2 = 2\alpha_1 p X^{p-1}, \quad \Delta_3 = 2\alpha_2 q Y^{q-1}$$
$$\Delta_4 = p(p-1)\alpha_1 X^{p-2}$$
$$\Delta_5 = q(q-1)\alpha_1 Y^{q-2}$$
$$\Delta_6 = \Omega^2 \left(1 + \beta_1 X^r + \beta_2 Y^s\right)$$

(4.6)

$$\Psi_1 = \frac{\partial^4 W}{\partial Y^4} - \frac{6W(i,j)}{\Delta Y^4}$$
$$\Psi_2 = \frac{\partial^4 W}{\partial X^2 \partial Y^2} - \frac{4W(i,j)}{\Delta X^2 \Delta Y^2}$$
$$\Psi_3 = \frac{\partial^4 W}{\partial X^4} - \frac{6W(i,j)}{\Delta X^4}$$
$$\Psi_4 = \frac{\partial^3 W}{\partial X^3}, \quad \Psi_5 = \frac{\partial^3 W}{\partial X \partial Y^2}$$
$$\Psi_6 = \frac{\partial^3 W}{\partial Y^3}, \quad \Psi_7 = \frac{\partial^3 W}{\partial X^2 \partial Y}$$
$$\Phi_1 = \frac{W(i+1,j) + W(i-1,j)}{\Delta X^2}$$
$$\Phi_2 = \frac{W(i,j+1) + W(i,j-1)}{\Delta Y^2}$$
$$\Phi_3 = \frac{W(i,j+2) - 4\{W(i,j+1) + W(i,j-1)\} + W(i,j-2)}{\Delta X^2}$$

(4.7)

Similarly, the discretized boundary condition along the four edge of the rectangular plates can be written as

### 4.1.2 For clamped edge

$$W(i,j) = \frac{\partial W}{\partial X} = \frac{W(i+1,j) - W(i,j)}{\Delta X} = 0 \text{ at } X = 0 \text{ and } X = 1$$

$$W(i,j) = \frac{\partial W}{\partial Y} = \frac{W(i,j+1) - W(i,j)}{\Delta Y} = 0 \text{ at } Y = 0 \text{ and } Y = 1 \quad (4.8)$$

### 4.1.3 for simply supported edge

$$W(i,j) = \frac{\partial^2 W}{\partial X^2} = \frac{W(i+1,j) + W(i-1,j) - 2W(i,j)}{\Delta X^2} = 0 \text{ at } X = 0 \text{ and } X = 1$$

$$W(i,j) = \frac{\partial^2 W}{\partial Y^2} = \frac{W(i,j+1) + W(i,j-1) - 2W(i,j)}{\Delta Y^2} = 0 \text{ at } Y = 0 \text{ and } Y = 1 \quad (4.9)$$

With using of the boundary condition in the converted governing equation of the plate we get the solution of this plate model.

# Chapter 5

# NUMERICAL RESULT AND DISCUSSION

## 5.1. NUMERICAL RESULT AND DISCUSSION

The numerical solution of equation (4.17) for each boundary condition has been obtained using the software MATLAB. Out of the various eigen values for each set of the value of the plate parameters, the lowest three eigen values have been retained as the first three natural frequencies. In this study there is one variation of non-homogeneity namely, linear i.e. $p = q = r = s = 1$ (LVN) and parabolic variation i.e. $p = q = r = s = 2$ considered for the illustration with the values of the other plate parameters.

The variation of the non-homogeneity parameters $\alpha_1$ and $\alpha_2 = -0.5(0.1)0.5$; and the density parameters $\beta_1$ and $\beta_2 = -0.5(0.1)0.5$; and the aspects ratio $a/b = 1$; and poison's ratio $\vartheta = 0.3$. To choose an appropriate number of grid points $(N, M)$, convergence study of frequency parameters $\Omega$ with increasing value of grid points $N = M = 8(2)22$, for homogeneous, 1D non-homogeneous and 2D non-homogeneous plate has been carried out for the different value plate parameters. For the evaluation of the frequency parameters $\Omega$ assumed equally spacing grid points.

Table 1 Convergence of frequency Parameters $\Omega$ with the grid points $(N, M)$ for homogeneous $(\alpha_1 = \beta_1 = \alpha_2 = \beta_2 = 0)$ square plate

| Mode → <br> No. of grid points | I | II | III | I | II | III |
|---|---|---|---|---|---|---|
| | | CCCC | | | CSCS | |
| $N = M = 8$ | 34.9911 | 71.0297 | 71.0301 | 27.5662 | 53.6742 | 65.0396 |
| $N = M = 10$ | 34.9815 | 72.2956 | 72.2956 | 27.9560 | 52.7323 | 68.1993 |
| $N = M = 12$ | **34.9815** | 72.3939 | 72.3939 | **27.9509** | 53.7431 | 68.3270 |
| $N = M = 14$ | - | **72.3989** | **72.3939** | - | **53.7431** | **68.3270** |
| $N = M = 16$ | - | - | - | - | - | - |
| $N = M = 18$ | - | - | - | - | - | - |
| $N = M = 20$ | - | - | - | - | - | - |
| $N = M = 22$ | - | - | - | - | - | - |

Table 1 shows that after taking $(\alpha_1 = \beta_1 = \alpha_2 = \beta_2 = 0)$ the plate model become homogeneous because the Young's modulus function become $E(X,Y) = E_0$ as well as mass density function become $\rho(X,Y) = \rho_0$. The first three modes of vibration for a particular set of the plate parameters where the maximum value of the grid points was required. From the table, it is clear that number of grid points required the same accuracy is in the order of the nature of the plate. It is found that for the same set of the values of the other plate parameters the frequency parameters for the CCCC edge is higher than CSCS edge for each and every mode.

Table 2 Convergence of frequency Parameters $\Omega$ with the grid points $(N, M)$ for homogeneous $(\alpha_1 = \beta_1 = 0.5, \alpha_2 = \beta_2 = 0)$ square $a/b = 1$ plate

| | Mode → No. of grid points ↓ | I | II CCCC | III | I | II CSCS | III |
|---|---|---|---|---|---|---|---|
| **LVN** | $N = M = 8$ | 34.3790 | 71.4567 | 71.3676 | 27.1498 | 53.8239 | 67.2525 |
| | $N = M = 10$ | 34.1039 | 72.2956 | 72.3956 | 27.9560 | 53.7323 | 68.1993 |
| | $N = M = 12$ | 35.1038 | 72.3100 | 73.3939 | **27.9560** | 54.6856 | 69.2100 |
| | $N = M = 14$ | **35.1038** | **72.3100** | **73.3939** | - | 54.6858 | **69.2100** |
| | $N = M = 16$ | - | - | - | - | **54.6858** | - |
| | $N = M = 18$ | - | - | - | - | - | - |
| | $N = M = 20$ | - | - | - | - | - | - |
| | $N = M = 22$ | - | - | - | - | - | - |
| **PVN** | $N = M = 8$ | 24.0596 | 28.1413 | 38.2386 | 23.7852 | 26.6687 | 33.5612 |
| | $N = M = 10$ | 24.0526 | 28.1513 | 35.5876 | 23.7800 | 26.6587 | 33.1521 |
| | $N = M = 12$ | 24.0425 | 28.1617 | 35.5261 | **23.7800** | 26.6566 | 33.1421 |
| | $N = M = 14$ | **24.0425** | 28.1915 | 35.1264 | - | 26.6512 | 32.8962 |
| | $N = M = 16$ | - | 28.1916 | 35.1265 | - | **26.6512** | 32.8742 |
| | $N = M = 18$ | - | **28.191** | 35.1265 | - | - | 32.8732 |
| | $N = M = 20$ | - | - | - | - | - | **32.8732** |
| | $N = M = 22$ | - | - | - | - | - | - |

Table 2 shows that when we take the different value of $\alpha_1, \alpha_2, \beta_1$ and $\beta_2$ other than zero, then non-homogeneity of this model come into existence because the Young's modulus function and the mass density function depends on the non-dimensional variable $X$ and $Y$. Since in the above table I had taken $(\alpha_1 = \beta_1 = 0.5, \alpha_2 = \beta_2 = 0)$ for considering the non-homogeneity the value of the frequency parameters $\Omega$ decreases from the previous value of the frequency parameters table for the mode shape I. It is found that the $\Omega$ for LVN as well as for PVN for the same set of the values of the other plate parameters is found in order of boundary condition CCCC > CSCS. It is found that the $\Omega$ for the CCCC boundary condition for the same set of the values of the other plate parameters is found in order of LVN > PVN.

Table 3 Convergence of frequency Parameters $\Omega$ with the grid points $(N, M)$ for homogeneous $(\alpha_1 = \beta_1 = \alpha_2 = \beta_2 = 0.5)$ square $a/b = 1$ plate

| | Mode → No. of grid points ↓ | I | II CCCC | III | I | II CSCS | III |
|---|---|---|---|---|---|---|---|
| **LVN** | $N = M = 8$ | 35.9958 | 72.8952 | 72.3676 | 28.9952 | 54.9853 | 68.5231 |
| | $N = M = 10$ | 35.9526 | 72.7625 | 72.3256 | 28.9862 | 54.9762 | 68.6542 |
| | $N = M = 12$ | 35.9214 | 72.2156 | 72.3210 | 28.9861 | 54.9642 | 69.5362 |
| | $N = M = 14$ | **35.9214** | **72.2156** | **72.3210** | **28.9861** | 54.8942 | 69.2489 |
| | $N = M = 16$ | - | - | - | - | **54.8942** | **69.2489** |
| | $N = M = 18$ | - | - | - | - | - | - |
| | $N = M = 20$ | - | - | - | - | - | - |
| | $N = M = 22$ | - | - | - | - | - | - |
| **PVN** | $N = M = 8$ | 36.6214 | 73.5082 | 73.5216 | 29.4098 | 55.4362 | 69.1398 |
| | $N = M = 10$ | 36.5642 | 73.4521 | 74.1520 | 29.4087 | 55.9030 | 69.9421 |
| | $N = M = 12$ | 36.5542 | 74.0920 | 74.1463 | 29.4092 | 55.1107 | 69.9511 |
| | $N = M = 14$ | **36.5542** | **74.0920** | 74.1205 | 29.4044 | 55.1102 | **69.5111** |
| | $N = M = 16$ | - | - | **74.1205** | **29.4044** | 55.1101 | - |
| | $N = M = 18$ | - | - | - | - | **55.1101** | - |
| | $N = M = 20$ | - | - | - | - | - | - |
| | $N = M = 22$ | - | - | - | - | - | - |

Table 3 shows that when we take the different value of $\alpha_1, \alpha_2, \beta_1$ and $\beta_2$ other than zero, then non-homogeneity of this model come into existence because the Young's modulus function and the mass density function depends on the non-dimensional variable $X$ and $Y$. Since in the above table I had taken $(\alpha_1 = \beta_1 = \alpha_2 = \beta_2 = 0.5)$ for considering the non-homogeneity the value of the frequency parameters $\Omega$ decreases from the previous value of the frequency parameters table for the mode shape I. It is found that the $\Omega$ for LVN as well as for PVN, for the same set of the values of the other plate parameters is found in order of boundary condition CCCC > CSCS. It is found that the $\Omega$ for the CCCC boundary condition for the same set of the values of the other plate parameters is found in order of PVN > LVN.

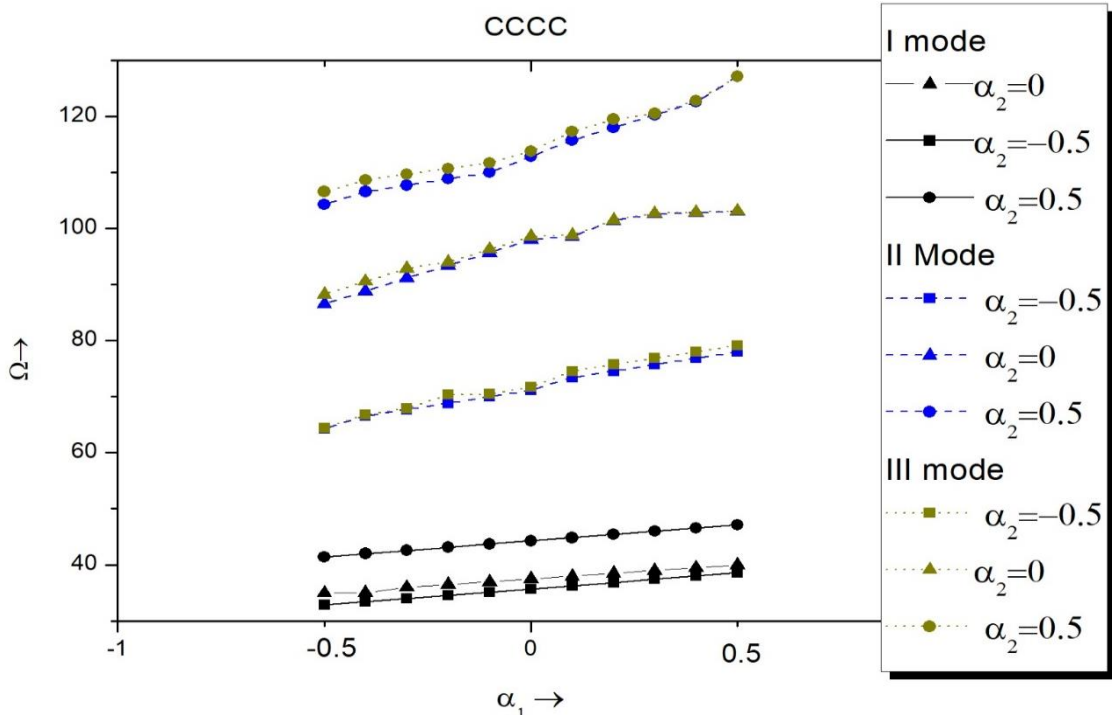

Figure 9 Frequency parameters $\Omega$ versus non-homogeneity parameters $\alpha_1$ for a square CCCC plate for $\beta_1 = \beta_2 = -0.5$, for the first, second and third mode respectively

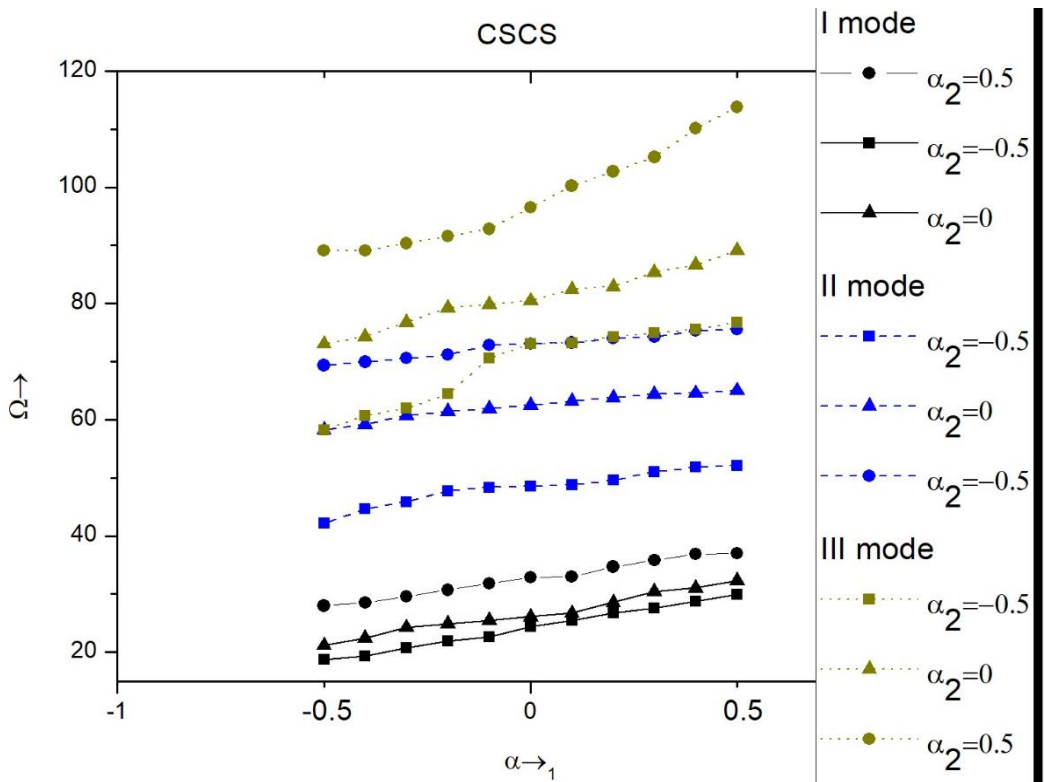

Figure 10 Frequency parameters $\Omega$ versus non-homogeneity parameters $\alpha_1$ for a square CSCS plate for $\beta_1 = \beta_2 = -0.5$, for the first, second and third mode respectively

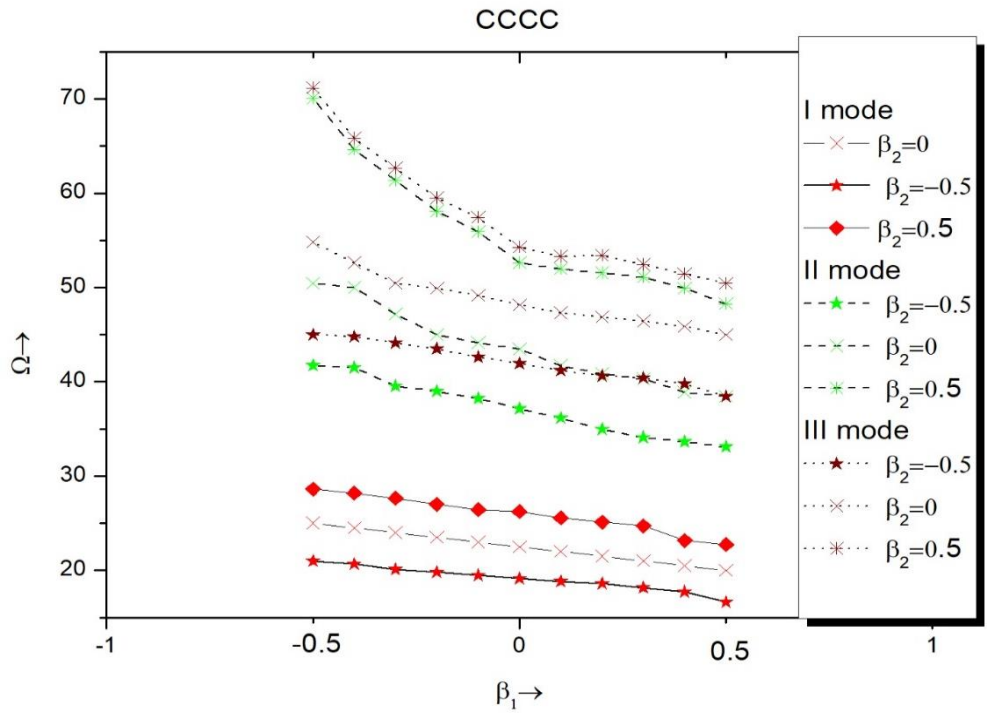

Figure 11 Frequency parameters $\Omega$ versus non-homogeneity parameters $\beta_1$ for a CCCC square $\alpha_1 = \alpha_2 = -0.5$, for the first, second and third mode respectively

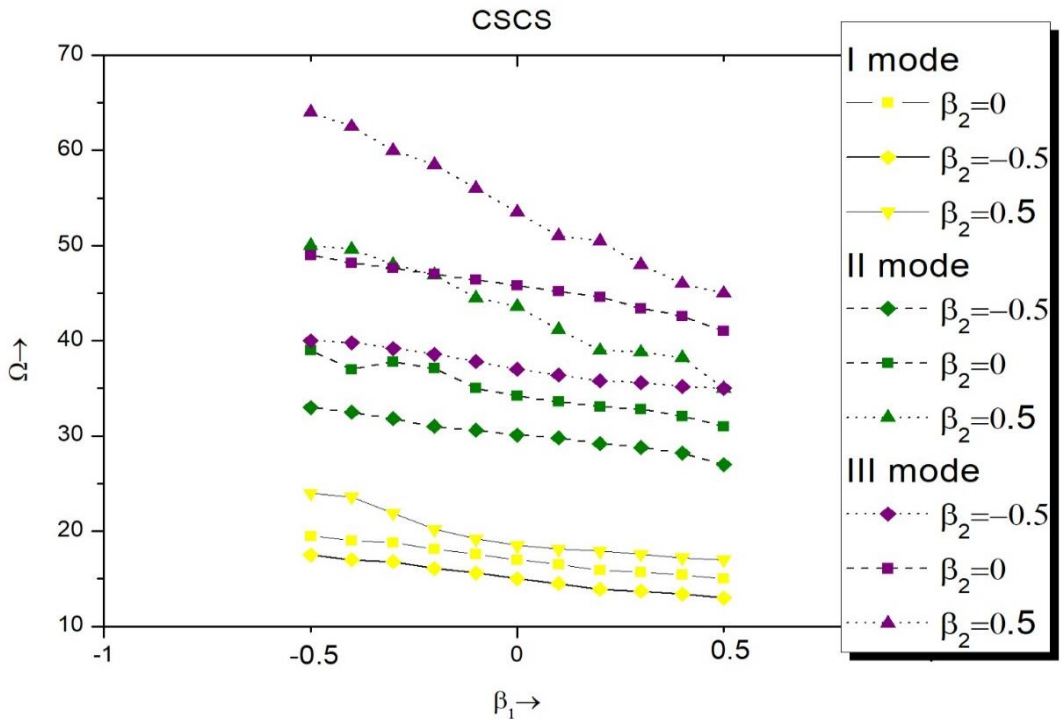

Figure 12. Frequency parameters $\Omega$ versus non-homogeneity parameters $\beta_1$ for a CSCS square $\alpha_1 = \alpha_2 = -0.5$, for the first, second and third mode respectively

Figure 9 and 10 shows the graphs for the frequency parameters $\Omega$ versus non homogeneity parameters $\alpha_1$ for the square plate with the fixed value of $\beta_1 = \beta_2 = -0.5$ and three different values of $\alpha_2 = -0.5, 0, 0.5$ for the first three mode of vibration for the two different boundary condition. It is found that the frequency parameters $\Omega$ increases as the stiffness of the plate along X-direction (i.e. $\alpha_1$) increases whatever the value of the $\alpha_2$. It further shows that frequency parameters $\Omega$ increases with stiffness of the plate along Y-direction (i.e. $\alpha_2$) changes from -0.5 to 0.5. The rate of increase of frequency parameters $\Omega$ with $\alpha_1$ increases with increasing the number of modes for the given boundary conditions CCCC > CSCS for the same set of the values of the other parameters. The effect of the $\alpha_2$ on the frequency parameters $\Omega$ is more pronounced for $\alpha_1 = -0.5$ as compared to $\alpha_1 = 0.5$.

Figure 11 and 12 shows that Frequency parameters $\Omega$ versus non-homogeneity parameters $\beta_1$ for a square $\alpha_1 = \alpha_2 = -0.5$, for the first, second and third mode respectively. The frequency parameters $\Omega$ decreases as the value of density parameters $\beta_1$ increases i.e. plate density becomes higher along X-direction whatever the value of $\beta_2$ for the both the boundary condition. Also the frequency parameters decreases due to increasing the value of $\beta_2$. It means the plate also become denser in Y- direction by increasing the value of $\beta_2$. The rate of decreasing the value of $\Omega$ with $\beta_1$ is higher in CCCC plate boundary condition as compare to CSCS plate boundary condition for the same set of the values of the other plate parameters.

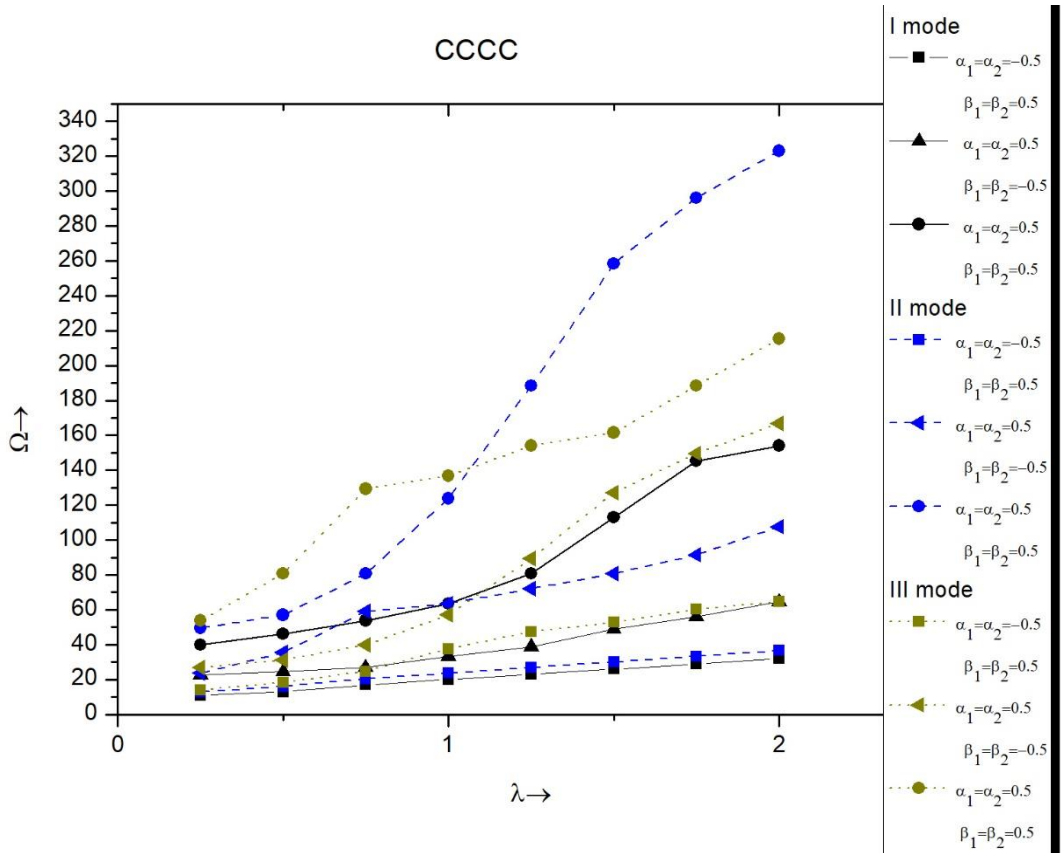

Figure 13 frequency parameters $\Omega$ versus aspect ratio for CCCC plate

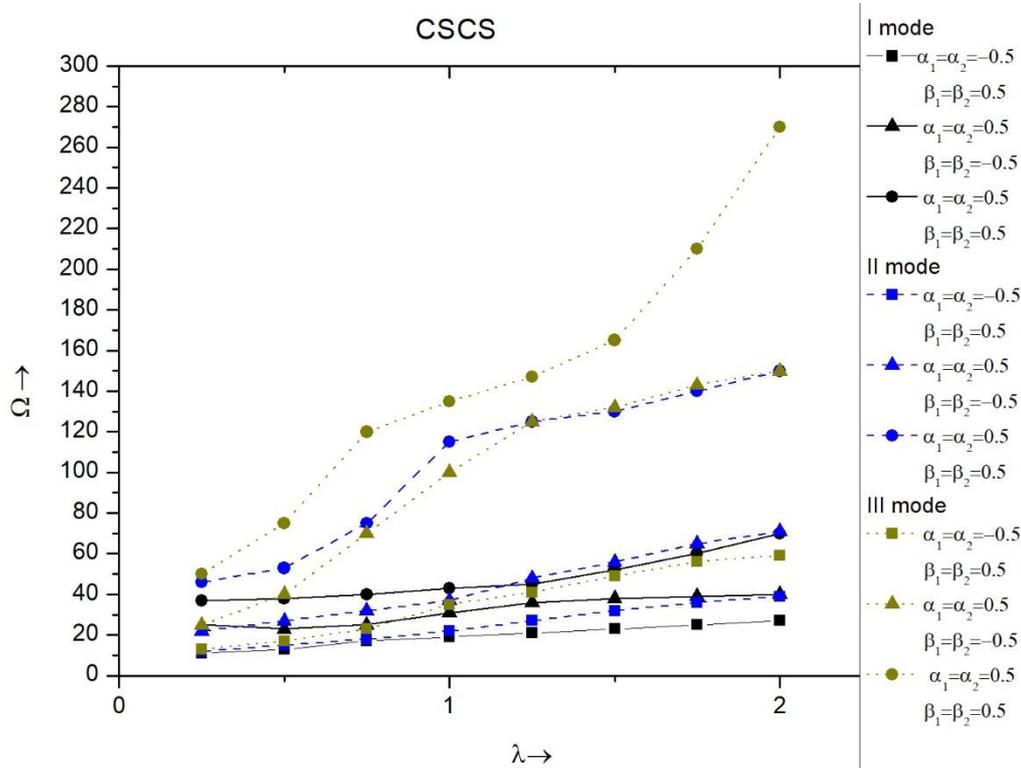

Figure 14 frequency parameters $\Omega$ versus aspect ratio for CSCS plate

Figure 13 and 14 shows the behaviour of the aspect ratio $\lambda$ on the frequency parameters $\Omega$ for $\alpha_1 = \alpha_2 = \pm 0.5$, $\beta_1 = \beta_2 = \pm 0.5$ for the two different boundary condition CCCC and CSCS plate for the first three mode of vibration. The graph for $\beta_1 = \beta_2 = \alpha_1 = \alpha_2 = -0.5$ has not be shown in the figure because it was overlapping with each other graphs. From the figure it has clear shown that the frequency parameters $\Omega$ increases with the increasing of the aspects ratio $\lambda$ for all the three modes of vibration. When the value of $\lambda$ becomes large for these boundary condition the plate behaves like a beam of length $a$, C/S at both the ends undergoes the anticlastic bending. Due to which the frequency parameters $\Omega$ increases. The effect is more prolonged when $\lambda \geq 1$ and in the order of the boundary condition CCCC > CSCS for the same set of the values of other plate parameters. And this effect also increases with increase in the number of modes.

### 5.1.4. Mode Shape

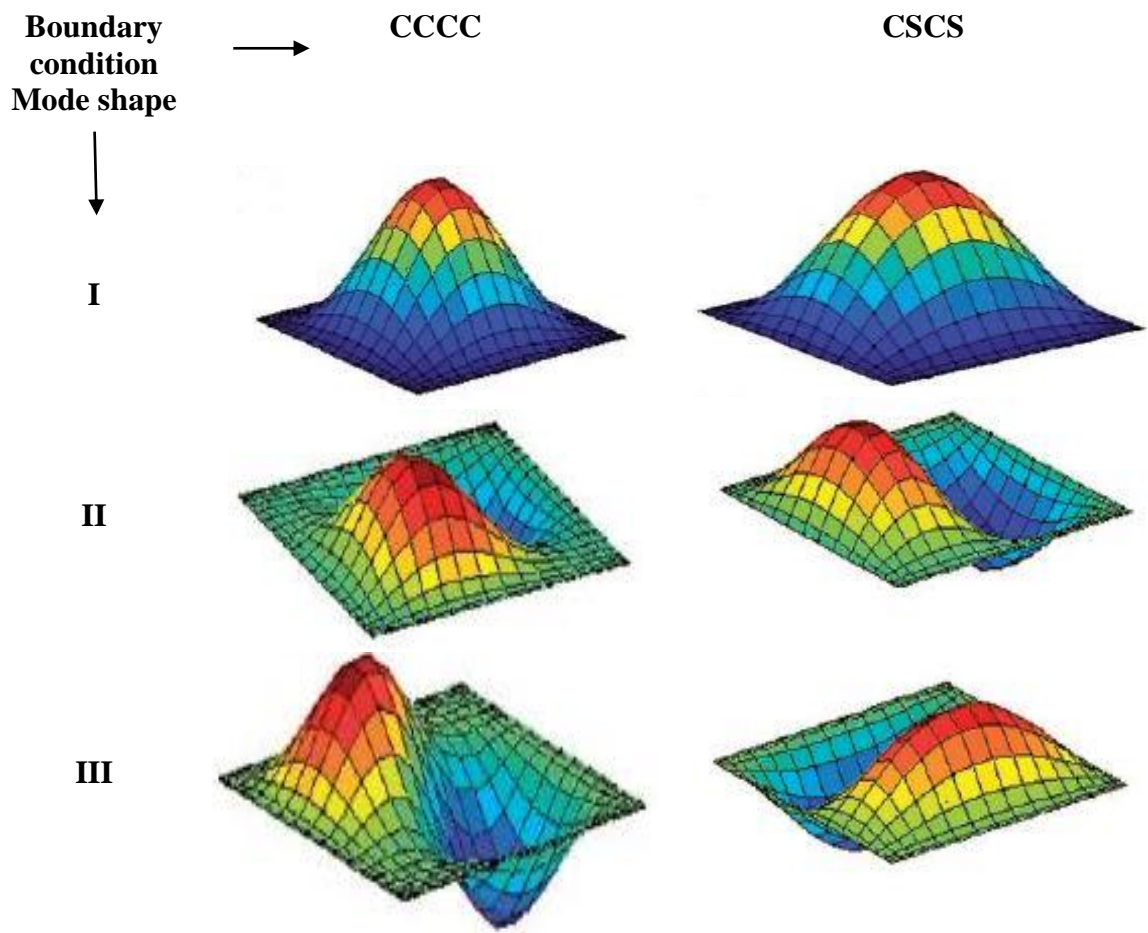

Figure 15 Mode shape

Figure 15 shows various modes of vibration a) CCCC and b) CSCS for $\beta_1 = \beta_2 = \alpha_1 = \alpha_2 = 0.5$ and $\lambda = 1$.

# Chapter 6

# CONCLUSION

## 6.1. Conclusion

The variation of the non-homogeneity of the plate material on the vibration characteristics of the rectangular plates has been analysed using the FDM technique method. For illustration the numerical result for linear variation for the non –homogeneity have been obtained for the first three modes of the vibration for the two different boundary condition CCCC and CSCS.

### 6.1.1 Observation
The following observation has been made:

(i) With the increase of non-homogeneity parameters of the plate, the non-dimensional frequency parameters increase when the density parameters of that plate material remain constant.

(ii) When the Young's modulus parameters kept constant, non-dimensional frequency parameters decreases with increasing of the density parameters.

(iii) For the both boundary condition non-dimensional frequency parameters increase with increase the value of aspects ratio $\lambda$ when the other parameters of the plate remain constant.

(iv) When there is no change in the other parameters of the plate the non- dimensional frequency parameters for the PVN is smaller as compared to LVN in non-homogeneity, when the plate density higher toward the edge $X=0$ and $Y=0$ as compared to $X=1$ and $Y=1$. When the change in order of density of the plate the results become the vice versa.

(v) When there is no change in the density parameters the value of the non-dimensional frequency parameters is higher for PVN as compared to LVN because, the plate is more stiff towards the edges $X=0$ and $Y=0$ as compared to the $X=1$ and $Y=1$. The results will be vice versa when the order of the stiffness changes from previous case.

### 6.1.2. Future Work
The above analysis is useful in the field of electronics and aerospace industry. Due to its non-homogeneity this paper can also give some idea in the field of metallurgy. This type of the non-homogeneity arises during the fiber reinforced plastic structure and in composite structure, which use fibers of different strength properties along two mutually perpendicular directions along the edge of the plate.

# REFERENCES

# References

1. Bardell NS (1991) Free vibration analysis of a flat plate using the hierarchical finite element method.  Journal of Sound and Vibration 151: 263–289.

2. Bert CW, Jang SK and Striz AW (1988) Two new approximate methods for analysing free vibration of structural components. AIAA Journal 26: 612–618.

3. Bhat RB, Laura PAA, Gutierrez RG and Cortinez VH (1990) Numerical experiment on the determination of natural frequencies of transverse vibrations of rectangular plates of non-uniform thickness. Journal of Sound and Vibration 139: 205–219.

4. Das AK and Mishra DM (1971) Free vibrations of an isotropic nonhomogeneous circular plate. AIAA Journal 9:963–964.

5. Gupta AK, Saini M, Singh S and Kumar R (2012) Forced vibrations of non-homogeneous rectangular plate of linearly varying thickness. Journal of Vibration and Control 1–9.

6. Gupta US, Lal R and Sharma S (2007) Vibration of non- homogeneous circular plates with variable thick- ness. Journal of Sound and Vibration 302(1–2): 1–17

7. Huang M, Ma XQ, Sakiyama T, Matsuda H and Morita C (2007) Free vibration analysis of rectangular plates with variable thickness and point support. Journal of Sound and Vibration 300: 435–452.

8. Jin GY, Su Z, Shi SX, Ye TG and Gao SA (2014b) Three- dimensional exact solution for the free vibration of arbitrarily thick functionally graded rectangular plates with general boundary conditions.  Composite Structures 108:565–577.

9. Kerboua Y, Lakis AA, Thomas M and Marcouiller L (2007) Hybrid  method  for  vibration analysis  of  rectangular plates. Nuclear Engineering Design 237: 791–801.

10. Lal R and Dhanpati (2007) Transverse vibrations of non- homogeneous orthotropic rectangular plates of variable thickness:  A  spline technique.  Journal  of  Sound  and Vibration 306: 203–214.

11. Lal R and Kumar Y (2012a) Boundary characteristic orthogonal polynomials in the study of transverse vibrations of non-homogeneous rectangular plates with bilinear thickness variation. Shock and Vibration 19: 349–364.

12. Lal R and Saini R (2015) Buckling and vibration analysis of non-homogeneous rectangular Kirchhoff plates resting on two-parameter foundation. Meccanica 50: 893–913.

13. Leissa AW (1973)  Free  vibrations  of  rectangular  plates. Journal of Sound and Vibration 31: 257–293.

14. Olszak W (1958) Non-homogeneity in elasticity and plasticity. Poland: Pergamon press.

15. Ramkumar RL, Chen PC and Sanders WJ (1987) Free vibration   solution   for   clamped   orthotropic   plates   using Lagrangian multiplier technique. AIAA   Journal   25:146–151.

16. Shu C and Richards BE (1992) Application of generalized differential quadrature to solve two dimensional incompressible Naiver-Stokes equations. International Journal of Numerical Methods in Fluids 15: 791–798.

17. Shu C (2000) Differential Quadrature and Its Application in Engineering. London: Springer Verlag.

18. Tomar JS, Gupta DC and Jain NC (1982a) Axisymmetric vibrations of an isotropic non-homogenous circular plate of linearly varying thickness. Journal of Sound and Vibration 85: 365–370.

19. Tomar JS, Gupta DC and Jain NC (1984) Free vibrations of an isotropic non-homogeneous infinite plate of parabolic- ally varying thickness. Indian Journal of Pure Applied Mathematics 15: 211–220.

20. Sakata T, Takahashi K and Bhat RB (1996) Natural frequencies of orthotropic rectangular plates obtained by iterative reduction of the partial differential equation. Journal of Sound and Vibration 189: 89–101.

21. Huang M, Ma XQ, Sakiyama T, Matsuda H and Morita C (2007) Free vibration analysis of rectangular plates with variable thickness and point support. Journal of Sound and Vibration 300: 435–452.

22. Kerboua Y, Lakis AA, Thomas M and Marcouiller L (2007) Hybrid method for vibration analysis of rectangular plates. Nuclear Engineering Design 237: 791–801.

23. Bhaskar K and Sivaram A (2008) Un-truncated infinite series superposition method for accurate flexural analysis of isotropic/orthotropic rectangular plates with arbitrary edge conditions. Composite Structures 83: 83–92.

24. Shi J, Chen W and Wang C (2009) Free vibration analysis of arbitrary shaped plates by boundary knot method. Acta Mechanica Solida Sinica 22: 328–336.

25. Shu C and Richards BE (1992) Application of generalized differential quadrature to solve two-dimensional incompressible Naiver-Stokes equations. International Journal of Numerical Methods in Fluids 15: 791–798.

26. Striz AG, Wang X and Bert CW (1995) Harmonic differential quadrature method and applications to analysis of structural components. Acta Mechanica 111: 85–94.

27. Shu C and Chew YT (1997) Fourier expansion-based differential quadrature and its application to Helmholtz eigenvalue problems. Communications Numerical Methods in Engineering 13: 643–653.

28. Liu GR and Wu TY (2001) Vibration analysis of beams using the generalized differential quadrature rule and domain decomposition. Journal of Sound and Vibration 246: 461–481.

29. K.S Virdi (2006) Finite difference method for nonlinear analysis of structures.

30. Chakraverty S, Jindal R and Agarwal VK (2005) Flexural vibration of non-homogeneous elliptic plates. Indian Journal of Engineering & Materials Sciences 12: 521–528.

31. Gutierrez RH, Laura PAA, Bambill DV, Jederlinic VA and Hodges DH (1998) Axisymmetric vibrations of solid circular and annular membranes with continuously varying density. Journal of Sound and Vibration 212(4): 611–622.

32. Dipak Kr. Maiti & P. K. Sinha. Bending and free vibration analysis of shear deformable laminated composite beams by finite element method. Composite Structures, 29 (1994): 421- 431.

32. Najafov AM, Sofiyev AH and Kuruoglu N (2014) Vibration analysis of nonhomogeneous orthotropic cylindrical shells including combined effect of shear deformation and rotary inertia. Meccanica 49: 2491–2502.

33. Tornabene F, Fantuzzi N, Viola E and Carrera E (2014) Static analysis of doubly-curved anisotropic shells and panels using CUF approach, differential geometry and differential quadrature method. Composite Structures 107: 675–697.

34. Sofiyev AH (2014) The influence of non-homogeneity on the frequency-amplitude characteristics of laminated orthotropic truncated conical shell. Composite Structures 107:334–345.

35. Jones RM (1999) Mechanics of composite materials, 2nd edn. Philadelphia: Taylor & Francis.